\documentclass[%
superscriptaddress,
 amsmath,amssymb,
 aps,
prl,
floatfix,
longbibliography
]{revtex4-1}
\usepackage[utf8]{inputenc}
\usepackage{graphicx}% Include figure files
\usepackage{dcolumn}% Align table columns on decimal point
\usepackage{bm}% bold math
\usepackage{hyperref}% add hypertext capabilities
\usepackage{amsmath,amsfonts,amssymb} 
\usepackage{pdfpages}
\usepackage{float}
\usepackage{braket}
\usepackage{tikz}
\usepackage{CJK}
\usepackage{siunitx}
\usepackage{xcolor}
\usepackage{tabularx}
\makeatletter
\AtBeginDocument{\let\LS@rot\@undefined}
\makeatother
\newcommand{\vF}{\ensuremath v_{\rm{F}} }
\newcommand{\dT}{\ensuremath d_\mathrm{t}}
\newcommand{\dB}{\ensuremath d_\mathrm{b}}
\newcommand{\dBLG}{\ensuremath d_\mathrm{g}}
\newcommand{\dCC}{\ensuremath d_\mathrm{CC}}
\newcommand{\dhBN}{\ensuremath d_\mathrm{hBN}}
\newcommand{\const}{\ensuremath \mbox{const.}}
\newcommand{\CT}{\ensuremath C_\mathrm{tg}}
\newcommand{\CB}{\ensuremath C_\mathrm{bg}}
\newcommand{\CBLG}{\ensuremath C_\mathrm{m}}
\newcommand{\EFT}{\ensuremath E_\mathrm{F1}(n_\mathrm{t})}
\newcommand{\EFB}{\ensuremath E_\mathrm{F2}(n_\mathrm{b})}
\newcommand{\VT}{\ensuremath V_\mathrm{tg}}
\newcommand{\VB}{\ensuremath V_\mathrm{bg}}

\newcommand{\DOST}{{\ensuremath {\cal D}_\mathrm{t}(E_\mathrm{F})}}
\newcommand{\DOSB}{{\ensuremath {\cal D}_\mathrm{b}(E_\mathrm{F})}}

\newcommand{\CQT}{{\ensuremath C_\mathrm{qt}}}
\newcommand{\CQB}{{\ensuremath C_\mathrm{qb}}}
\newcommand{\KT}{{\ensuremath \mathbf{K}_\mathrm{t}}}
\newcommand{\KB}{{\ensuremath \mathbf{K}_\mathrm{b}}}
\newcommand{\dG}{{\ensuremath dG/d\VT}}
\newcommand{\nT}{{\ensuremath n_\mathrm{t}}}
\newcommand{\nB}{{\ensuremath n_\mathrm{b}}}
\newcommand{\gT}{{\ensuremath g_\mathrm{t}}}
\newcommand{\gB}{{\ensuremath g_\mathrm{b}}}

\newcommand{\sigmaT}{\ensuremath \sigma_\mathrm{T}}

\newcommand{\EBLG}{\ensuremath E_\mathrm{BLG}}

\newcommand{\epsilonG}{\ensuremath \epsilon_\mathrm{g}}
\newcommand{\epsilonhBN}{\ensuremath \epsilon_\mathrm{hBN}}
\newcommand{\epsBLG}{\ensuremath \epsilon_\mathrm{BLG} \epsilon_0}

\newcommand{\ET}{\ensuremath E_\mathrm{t}}
\newcommand{\EB}{\ensuremath E_\mathrm{b}}
\newcommand{\Vtop}{\ensuremath V_\mathrm{t}}
\newcommand{\Vbottom}{\ensuremath V_\mathrm{b}}
\newcommand{\sign}{\text{sgn}}

\begin{document}
%\preprint{APS/123-QED}
\title{The Electronic Thickness of Graphene}% Force line breaks with \\

\author{Peter Rickhaus}
%\email{peterri@phys.ethz.ch}
\affiliation{Solid State Physics Laboratory, ETH Zürich,~CH-8093~Zürich, Switzerland}
\author{Ming-Hao Liu}
\email{minghao.liu@phys.ncku.edu.tw}
\affiliation{Department of Physics, National Cheng Kung University, Tainan 70101, Taiwan}
\author{Marcin Kurpas}
\affiliation{Institute of Physics, University of Silesia in Katowice, 41-500 Chorzów, Poland}
\author{Annika Kurzmann}
\author{Yongjin Lee}
\author{Hiske Overweg}
\author{Marius Eich}
\author{Riccardo Pisoni}
\affiliation{Solid State Physics Laboratory, ETH Zürich,~CH-8093~Zürich, Switzerland}
\author{Takashi Tamaguchi}
\author{Kenji Wantanabe}
\affiliation{National Institute for Material Science, 1-1 Namiki, Tsukuba 305-0044, Japan}
\author{Klaus Richter}
\affiliation{Institute for Theoretical Physics, University of Regensburg, D-93040 Regensburg, Germany}
\author{Klaus Ensslin}
\author{Thomas Ihn}
\affiliation{%
Solid State Physics Laboratory, ETH Zürich,~CH-8093~Zürich, Switzerland}

\date{\today}% It is always \today, today,

\begin{abstract}
The van-der-Waals stacking technique enables the fabrication of heterostructures, where two conducting layers are atomically close. In this case, the finite layer thickness  matters for the interlayer electrostatic coupling. Here we investigate the electrostatic coupling of two graphene layers, twisted by $\theta=\SI{22}{\degree}$ such that the layers are decoupled by the huge momentum mismatch between the K and K' points of the two layers. 
We observe a splitting of the zero-density lines of the two layers with increasing interlayer energy difference. This splitting is given by the ratio of single-layer quantum capacitance over interlayer capacitance $\CBLG$ and is therefore suited to extract $\CBLG$. We explain the large observed value of $\CBLG$ by considering the finite dielectric thickness $\dBLG$ of each graphene layer and determine $\dBLG\approx\SI{2.6}{\angstrom}$. In a second experiment we map out the entire density range with a Fabry-Pérot resonator. We can precisely measure the Fermi-wavelength $\lambda$ in each layer, showing that the layers are decoupled. We find that $\lambda$  exceeds $\SI{600}{nm}$ at the lowest densities and can differ by an order of magnitude between the upper and lower layer. These findings are reproduced using tight-binding calculations.
\end{abstract}
%\begin{tocentry}
%\includegraphics[]{figTOC.pdf}
%\end{tocentry}
\maketitle

%%%%

%The electronic quality of  has improved drastically in the past decade and therefore the long-wavelength regime has become accessible for quantum transport experiments. This has led to a variety of electron-optics experiments in graphene \cite{Young2009,Taychatanapat2013,Rickhaus2013,Lee2015,Rickhaus2015,Eich2018}.
The van-der-Waals stacking technique allows scientists to bring two conductive crystalline layers into atomically close proximity \cite{Novoselov2016a}.
This has been exploited in a variety of experiments, including the formation of layer polarized, counter-propagating Landau levels\cite{Sanchez-Yamagishi2017}
 and experiments that build on strong capacitive coupling such as Coulomb-drag measurements \cite{Gorbachev2012} or interlayer exciton condensation\cite{Liu2017,Liu2018}.
 
\begin{figure}
 	\centering
 	\includegraphics[width=0.53\textwidth]{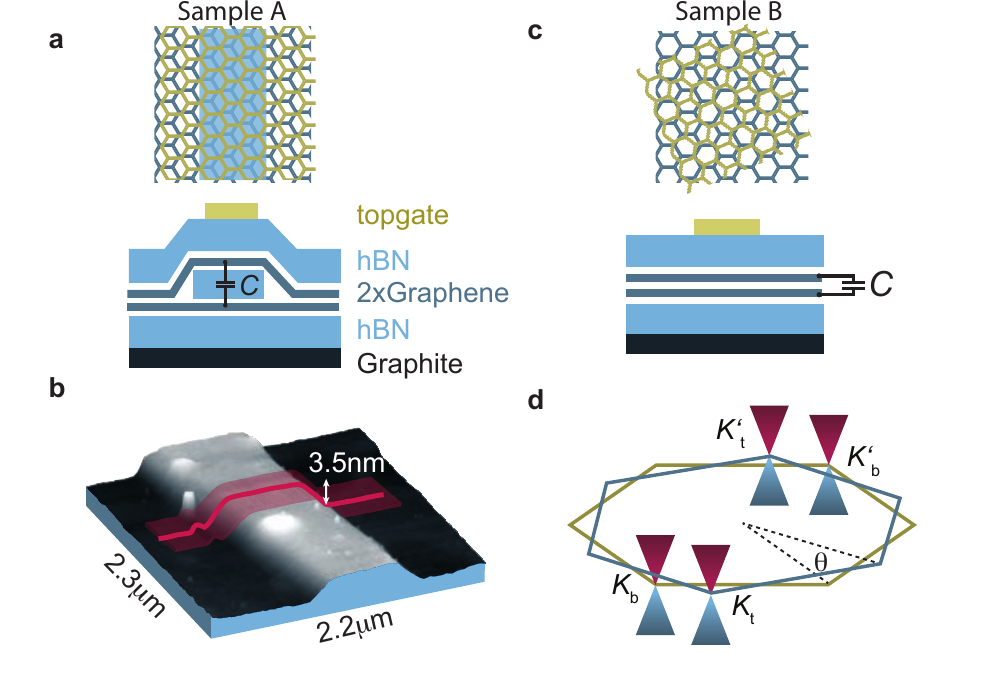}
 	\caption{a) Top-view and side-view of two aligned layers of graphene that are decoupled in the middle (blue part) by a thin intermediate layer of hBN. A graphite back gate and a local top gate allow to control the density and thereby the carrier wavelength in the upper- and lower layer individually.
 		b) Using atomic force microscopy we measured the encapsulated hBN layer to be 3.5nm thick (sample A).
 		c) Alternatively, the decoupling wavefunctions can be achieved by twisting two graphene layers (sample B).
 		d) For large twist angles, the valleys in the upper/lower layer ($\KT$, $\KB$) are separated by a large momentum, leading to an effective electronic decoupling of the layers. 
 	}\label{fig:concept}
 \end{figure}

There are two main approaches of how to bring two conductive layers in close proximity, while suppressing an overlap of the layer wavefunctions: One approach introduces a thin layer of hexagonal Boron-Nitride (hBN) (see e.g. \cite{Gorbachev2012,Greenaway2015,Liu2017}) as depicted in Fig. \ref{fig:concept}a,b, and the other twists the layers by a large angle ($\theta>5^\circ$) \cite{Lucian2011,Rozhkov2016,Sanchez2012,Sanchez-Yamagishi2017}; see Fig. \ref{fig:concept}c,d. In the former case, decoupling is achieved by spatial separation. In the latter case, the layers are ultimately close, but they remain decoupled due to a large momentum mismatch $(\KT-\KB)$ between the upper and lower layer (Fig. \ref{fig:concept}d). Experimental signatures of decoupling are an increased interlayer resistance\cite{Chari2016,Ribeiro-Palau2018} and layer-polarized Landau-levels at large magnetic fields \cite{Sanchez2012,Sanchez-Yamagishi2017}. 
%But even though the layers are electronically decoupled, the electrostatic coupling between them is very strong. For the capacitance between the layers, the finite dielectric thickness of graphene becomes crucial.

 In this work we perform quantum transport experiments  to monitor precisely the coupling, coherence and tunability of two graphene layers which are in close proximity to each other. In one device we separate the two layers by a thin layer of hBN with thickness $d=\SI{3.5}{nm}$  (sample A) and in the other device we twist the layers by $\SI{22}{\degree}$ in order to decouple them (sample B).
 
 In the first experiment we observe a splitting of the charge neutrality points of the two layers in the parameter plane of top- and back gate voltage ($\VT$, $\VB$). By analyzing the splitting we extract a geometric capacitance $\CBLG$ between the  graphene layers. For sample A we obtain the expected value given the thickness and dielectric constant of the intermediate hBN layer. However, for sample B,
 $\CBLG$ is three times larger than the geometric capacitance between two ideal capacitor plates, separated by the interlayer distance between carbon atoms $\dCC=\SI{3.4}{\angstrom}$ assuming vacuum in-between. We argue that, due to the finite electronic thickness of graphene, the plates of the capacitor are effectively closer than $\dCC$ leading to the enhanced $\CBLG$. We find good agreement with a capacitive model where we take the electronic thickness of graphene into account. 
  
In the second experiment on sample B we use a gate-defined Fabry-Pérot cavity to monitor the layer densities, coherence and interlayer coupling of wavefunctions. The cavities are formed by gate-defined p-n junctions which act as semi-transparent lateral "mirrors" of the interferometer \cite{Liang2001, Cheianov2006, Young2009, Rickhaus2013, Varlet2014}. Either only one or both layers can be tuned to the bipolar p-n-p regime. In both layers we observe the lowest energy Fabry-Pérot mode, corresponding to $\lambda=\SI{600}{nm}$, while the wavelength in the other cavity can be shorter by a factor of 10. We model the observed interference pattern using tight-binding calculations  assuming completely decoupled layers.
This second experiment confirms the assumed electronic decoupling and, for arbitrary gate voltages, the electrostatic model that considers thick graphene.

%The spacing of the Fabry-Pérot interferences allows to extract an energy scale, i.e. the tunnel coupling between the layers must be much smaller than the spacing of the Fabry-Pérot energy levels, i.e. $\ll\SI{1}{meV}$

\paragraph{Methods}

In order to achieve ballistic transport, we encapsulate\cite{Wang2013} either twisted bilayer graphene (sample B) or graphene-($\SI{3.5}{nm}$ hBN)-graphene between hBN layers (sample A) and use a graphite bottom gate \cite{Zibrov2017a,Overweg2017a}. The alignment of the graphene layers is controlled by the method described in Refs. \cite{Kim2016,Kim2017}, and we employed twist angles (between the graphene layers) $\theta\approx\SI{0}{\degree}$ for sample A and $\theta\approx\SI{22}{\degree}$ for sample B.  The thickness of the top, bottom and intermediate hBN layers is determined by atomic force microscopy (AFM).
Electrical one-dimensional contacts are achieved by reactive ion etching and evaporation of Cr/Au. Top gates of size $\SI{320}{nm}$ and $\SI{190}{nm}$ are defined by electron beam lithography. By adjusting the top gate $\VT$ and back gate voltage $\VB$, a Fabry-Pérot cavitiy can be formed below the top gate.  
Two-terminal linear conductance measurements are performed using a low-frequency lock-in technique ($\SI{177}{Hz}$) at the temperature $T=\SI{1.5}{K}$.

\begin{figure*}
	\centering
	\includegraphics[width=1\textwidth]{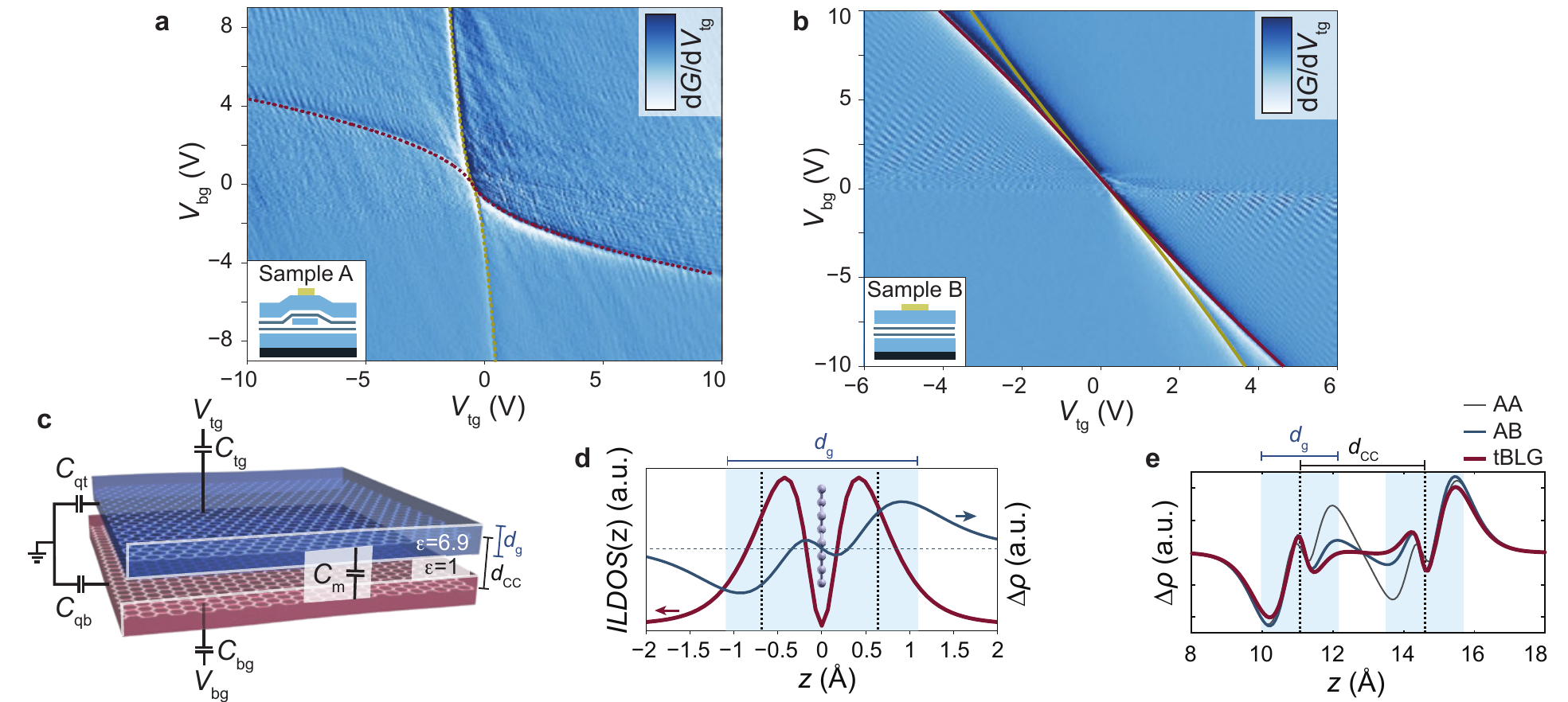}
	\caption{Numerical derivative of the two-terminal conductance, $\dG(\VT,\VB)$ for a device where the graphene layers are a), separated by a thin hBN layer (sample A) or
	b), in atomic vicinity, but twisted by a large angle (sample B). Zero-density lines in the upper (yellow) and lower (red) graphene layer are obtained from numerical calculations.
	c) Schematic electrostatic configuration of sample B.
	d)  Calculated integrated local density of states $ILDOS(z)$ of $p_z$-like orbital of carbon atoms in graphene (red) and the induced charge density $\Delta \rho (z):=\rho (0) - \rho (E_z)$ per carbon atom under an external electric field 
	$E_z$. The geometry of graphene is shown on the background picture. The graphene sheet is placed at $z=0$ and extends in the \textit{xy} plane. Positions of black dotted lines mark the effective thickness of graphene calculated from the expectation value of the position operator 
	$\left\langle z\right\rangle = \SI{0.66}{\angstrom}$. Blue shaded region shows the dielectric thickness of graphene extracted from the dielectric permitivity \cite{fang2016}.
	e) Comparison of $\Delta \rho (z)$ for bilayer graphene in AA stacking configuration (gray line), AB bernal (blue line) and twisted BLG (red line). The position of graphene layers is marked by vertical dashed lines and blue shaded regions depict the dielectric thickness of single-layer graphene.}
	\label{fig:splitting}
\end{figure*}

\paragraph{Results}

The numerical conductance $dG/d\VT$ as a function of $\VT$ and $\VB$ is shown in Fig. \ref{fig:splitting}a for sample A and Fig. \ref{fig:splitting}b for sample B. In both cases two pronounced curved lines are observed, corresponding to a dip in the conductance $G$. The lines cross at zero gate voltages and the splitting between these lines increases with increasing difference in $\VT$ and $\VB$. One line (following the yellow dashed line) is affected more strongly by the top gate voltage and therefore corresponds to the condition for charge neutrality in the upper graphene layer, whereas the other line (red dashed) indicates charge neutrality in the lower layer. 

From electrostatic considerations we find that the zero-density condition can be expressed as (details are given in the Supplemental Material)
\begin{eqnarray*}
\left.\frac{\partial\VT}{\partial\VB}\right|_{\nB=0} & \approx & -\frac{\CB}{\CT}\left(1+\frac{\CQT}{\CBLG}\right), \label{main1}\\
\left.\frac{\partial\VB}{\partial\VT}\right|_{\nT=0} & \approx & -\frac{\CT}{\CB}\left(1 + \frac{\CQB}{\CBLG}\right), \label{main2}
\end{eqnarray*}
where $\CB$  ($\CT$) is the geometric capacitance of the bottom (top) graphene to the bottom- (top-) gate (see Fig. \ref{fig:splitting}c) and the density in the bottom (top) graphene layer is $\nB$ ($\nT$). The capacitance measured between the two graphene plates is $\CBLG$. The quantum capacitance $\CQT=e^2\DOST$ of the top layer is proportional to the density of states at the Fermi energy in the top layer (the analogue relation holds for the bottom layer). For a single-sheet of graphene, the slope of the zero-density line in a ($\VT,\VB$)-map is given by the ratio  $-\CB/\CT$ (prefactor in the above equations). For the two-layer system, the deviations from linearity of the constant density line are governed by the ratio between quantum capacitance and 
$\CBLG$, respectively. Therefore the splitting is smaller in sample B where $\CBLG$ is large as compared to sample A, where $\CBLG$ is smaller.

 Analytical formulas for the zero density lines (i.e. $\VB (\VT)|_{\nT=0} 
$ and $\VB (\VT)|_{\nB=0}$ ) can be calculated using the ideal density of states of defect-free graphene and are depicted in Fig. \ref{fig:splitting}a and b for the different electrostatic configuration (i.e. with or without hBN between the graphene sheets). The formulas and details of the calculation are given in the Supplemental Information. Fitting these curves to the data allows us to extract $\CBLG$ which is the only free fitting parameter. The other capacitances in the problem  are given by the thickness of the top/bottom hBN, i.e. $\CT=\epsilonhBN/\dT$ with $\epsilonhBN=3.3\epsilon_0$. A discussion for the precision of this method is given in the Supplemental Material. \ref{app:accuracy}

%To obtain a value for the interlayer capacitance, we calculate the layer densities $\nT$ and $\nB$ for given $\VT$, $\VB$ and $\CT$, $\CB$ and identify lines where either $\nT=0$ or $\nB=0$. $\CT$ and $\CB$ are given by the thickness of the top/bottom hBN, i.e. $\CT=\epsilonhBN/\dT$ with $\epsilonhBN=3.3\epsilon_0$. We assume that the dispersion relation remains linear, such that the carrier density formulas \cite{Liu2013} derived for single-layer graphene with quantum capacitance \cite{Luryi1988,Fang2008} taken into account can be readily applied. However, the extremely thin spacing between the two graphene layers leads to significant electrostatic coupling. Effectively, the channel potential of the top layer plays the role as a gate for the bottom layer, and vice versa, requiring an iterative process for the solution to the carrier densities. Once the process converges for the layer densities we can identify lines of zero density in a $(\VT,\VB)$-map and adjust $\CBLG$ to fit our data. 

For sample A we obtain an interlayer capacitance of $\CBLG=\SI{0.81}{\mu Fcm^{-2}}$ which corresponds to the expected value for a plate separation of  $d=\SI{3.5}{nm}$ and the hBN dielectric constant of $\epsilonhBN=3.3\epsilon_0$.
For sample B, we determine a large interlayer capacitance $\CBLG=7.5\pm\SI{0.7}{\mu Fcm^{-2}}$. This value is three times larger than the capacitance between two thin plates, separated by vacuum and an interlayer distance of $\dCC=\SI{3.4}{\angstrom}$ which is the expected distance between two graphene layers \cite{Huang2006,Haigh2012}. 
Consistent with our findings, large interlayer capacitance values have been reported in Ref.\cite{Sanchez2012} in large perpendicular magnetic fields (quantum Hall regime) with a capacitance model that is only valid for $\nT=-\nB$. A detailed explanation for the large value of $\CBLG$ has not been given so far. 

To understand the origin of such a large effective interlayer capacitance we need to take into account the finite thickness of graphene as this reduces the effective distance between the capacitor plates, leading to an enhanced interlayer capacitance. Therefore we have estimated the extent of the $p_z$ orbitals of carbon atoms in graphene from first principles calculations (details are given in the Supplemental Material). We calculated the integrated local density of states profile $\rho(z)$ of single-layer graphene in the energy range $E \in [-3, 3]$\,eV from the charge neutrality point  at $E_c=0$\,eV. In this energy range the bands are of pure $p_z$ orbital character without contributions from the $s$-, $p_x$- and $p_y$-like orbitals (not shown here). The calculated integrated local density of states $ILDOS(z)$ as a function of distance from the center of the carbon atom is shown in Fig. \ref{fig:splitting}d. From the charge distribution we then calculated the expectation value of the position operator  $\langle z\rangle$ for one lobe of $p_z$ orbital (positive $z$). The values are shown as black dashed lines in the figure. 
Since there is a substantial amount of charge at $|z| > \langle z\rangle$, we have to take into account the induced charge density $\Delta \rho=\rho(E=0)-\rho(E)$ in an external electric field $E$ which determines the
\textit{dielectric thickness} of graphene \cite{fang2016}, defined as the distance from the center of carbon atoms to the point at which the dielectric constant of graphene $\epsilon = 6.9\epsilon_0$ decays to the vacuum permitivity. The dielectric thickness is the relevant quantity if considering a single-layer of graphene to be a nanocapacitor on its own. The dielectric thickness of graphene $\dBLG$ is indicated by the blue shaded region in Fig. \ref{fig:splitting}d with values according to Ref.\cite{fang2016}. 

In order to check if tBLG displays a qualitatively different electrostatic behavior than AA and AB-stacked BLG we performed first-principles calculations of twisted bilayer graphene with a twist angle 22$^\circ$ (details of computations are given in Supplemental Material). 
In Fig. \ref{fig:splitting}e we show the comparison of the induced charge density $\Delta \rho(z):=\rho (0) - \rho (E_{\text{z}})$ for tBLG, AA BLG and AB BLG under an external electric field $E_{\text{z}}=1\,\text{V/nm}$ perpendicular to the BLG lattice. 
The interlayer distance of AA and AB BLGs was set to $3.51\,\text{\AA}$ to fit the average distance between tBLG layers.
Nevertheless, the results are representative and insensitive to small deviations of interlayer distance from the optimized value or to the choice of the dispersive correction due to vdW forces (see Supplemental Material).
One can see that the responses of the different BLGs to the external electric field $E_{\text{z}}$ are almost the same on the outer side of the BLG, while they are very different in the interlayer region. For $z=13 \pm \SI{0.7}{\angstrom}$ we observe a flattening of $\Delta \rho(z)$ in case of  tBLG  compared to AA and AB BLG. Within this region the amplitude of $\Delta \rho(z)$ for tBGL is 15 times smaller than for AB BLG and 50 times smaller for AA BLG, demonstrating a qualitatively different electrostatic picture.

These calculations motivate a simplified capacitance model where the measured capacitance $\CBLG$ (between the center of charge of each layer), contains two dielectric materials coupled in series: Graphene with $\epsilonG = 6.9\epsilon_0$ \cite{fang2016} and thickness $\dBLG$ and an interlayer region of vacuum with thickness $d_\mathrm{inter}=\dCC-\dBLG$ and a dielectric constant of vacuum. Therefore 
$1/\CBLG=\dBLG/\epsilonG+(\dCC-\dBLG)/\epsilon_0$.
With $\dCC=\SI{3.4}{\angstrom}$ and the measured capacitance we determine a dielectric thickness of $\dBLG=2.6\pm\SI{0.2}{\angstrom}$ from our measurements which is in agreement with  theoretical predictions in single-layer graphene exhibiting 
$\SI{2.4}{\angstrom}$ \cite{fang2016}.  Using a similar model for the hBN device with $1/\CBLG=\dBLG/\epsilonG+\dhBN/\epsilonhBN$ we find $\dhBN=\SI{35}{\angstrom}$, in excellent agreement with the thickness measured with the AFM. However the correction by the thickness of graphene ($\approx\SI{1}{\angstrom}$) in this case is of the order of the measurement accuracy of our AFM.

\begin{figure*}
	\centering
	\includegraphics[width=1\textwidth]{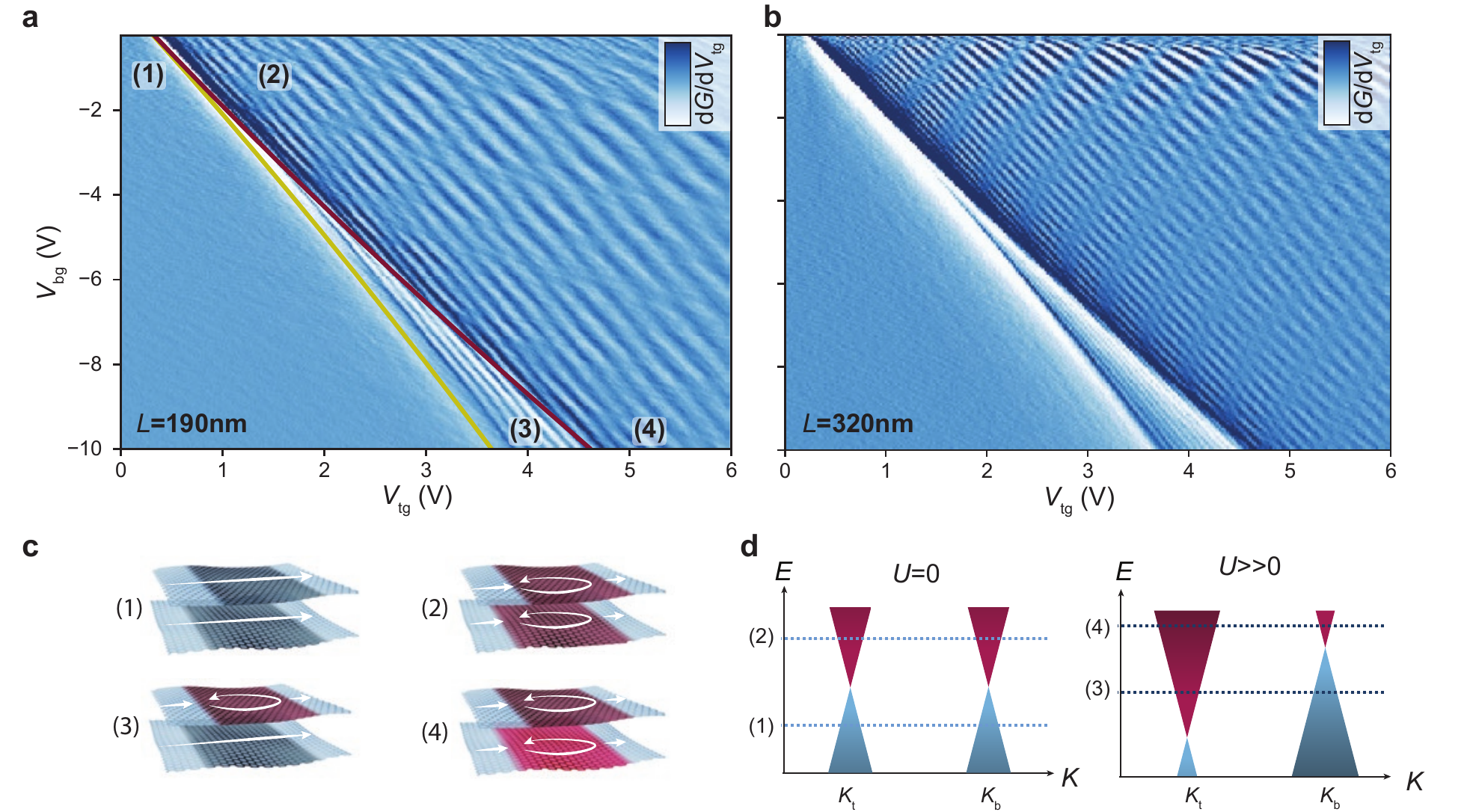}
	\caption{Differential conductance $\dG(\VT,\VB)$ for gates of length a) $L=\SI{190}{nm}$ and b) $\SI{320}{nm}$ for sample B. 
	c) Sketches of the local density in the two Fabry-Pérot layers. Blue regions are p-doped and red regions n-doped. The sketches (1)-(4) show different gating configurations, marked correspondingly  in a).
	d) As the difference in gate voltages increases, the energies in the top/bottom layers will shift by the interlayer energy difference $U$.}
	\label{fig:FPdata}
\end{figure*}

In the next step we use a Fabry-Pérot interferometer to measure the layer density of sample B for arbitrary gate voltages and  compare the results to tight-binding simulations based on an elaborate electrostatic model. 
The analysis of the Fabry-Pérot resonance pattern will allow us to determine the Fermi wavelength in the individual layers and will reveal that the graphene layers are indeed electronically decoupled. In Figs. \ref{fig:FPdata}a and b we show $\dG$ for top gates, sized $L=\SI{190}{nm}$ and $L=\SI{320}{nm}$, respectively. For both cases, the cavity width $W\gg L$. The zero-density lines are depicted in yellow for the top- and dark red for the bottom layer. 

The Fabry-Pérot resonator exhibits a  pattern that can be qualitatively understood by considering the layer densities in the regions underneath and outside the top gate, as depicted in Fig. \ref{fig:FPdata}c.
The density in the single-gated outer regions is affected only by $\VB$.  Since $\VB<0$, the outer regions are p-doped (blue colored).  For small voltages, labeled (1) and (2) in Figs. \ref{fig:FPdata}a and c, the density of each of the two layers is comparable, i.e. there is only a small energy difference $U$ between the two layers (see Fig. \ref{fig:FPdata}d). A p-n-p cavity below the top gate is formed for a sufficiently positive top gate voltage (2) in both layers. Given a large energy difference between the layers, it becomes possible to create a p-n-p cavity in only one layer (3) or also in both (4).

 As soon as a p-n-p cavity is formed, the conductance is modulated by standing waves, leading to the observed resonance pattern in Fig. \ref{fig:FPdata}a and b. In the inner region (3), only one set of Fabry-Pérot resonances, related to zero density in the upper layer, is observed. For densities beyond the zero density line of the lower layer (dark red line in Fig. \ref{fig:FPdata}a), a more complex resonance pattern appears. 

The resonance pattern is determined by the Fabry-Pérot condition, where the j-th resonance is $j=2L/\lambda_{\rm{F}}=k_{\rm{F}}L/\pi$ with $L$ the cavity size and $\lambda_{\rm{F}}$ the Fermi wavelength. Note that $k_{\rm{F}}=\sqrt{\pi n}$ is given by the density in the top/bottom layer. As expected, we observe a finer spacing of the resonance pattern for the larger cavity (Fig. \ref{fig:FPdata}b with $L=\SI{320}{nm}$) as compared to the smaller cavity (Fig. \ref{fig:FPdata}a with $L=\SI{190}{nm}$). In the region between the zero-density lines, six resonances are observed at large $U$ for $L=\SI{190}{nm}$ and even ten resonances for $L=\SI{320}{nm}$, i.e. it is possible to fill ten modes in the upper resonator while there is still no cavity formed in the lower layer. By assuming that $L$ is given by the lithographic size it follows that  $\lambda_{\rm{F,bottom}}=\SI{640}{nm}$ and $\lambda_{\rm{F,top}}=\SI{64}{nm}$ once the first mode fits into the cavity in the bottom layer at large $U$. Therefore, the wavelength can differ by an order of magnitude between two graphene layers despite the fact that those layers are atomically close.

In the measurement, especially for the larger cavity (Fig. \ref{fig:FPdata}b) it can also be seen that the oscillation amplitude is largest for either small values of $\VB$ or close to the zero density lines. Under these conditions, either the graphene part tuned only by $\VB$ or cavity below the topgate are close to zero density and therefore the density profile along the junction is especially flat, leading to a smooth transition between the cavity and the outer region. The enhanced oscillation amplitude can be understood by considering that smooth p-n interfaces act as strong angular filters \cite{Cheianov2006, Rickhaus2013}.

\begin{figure*}
	\centering
	\includegraphics[width=1\textwidth]{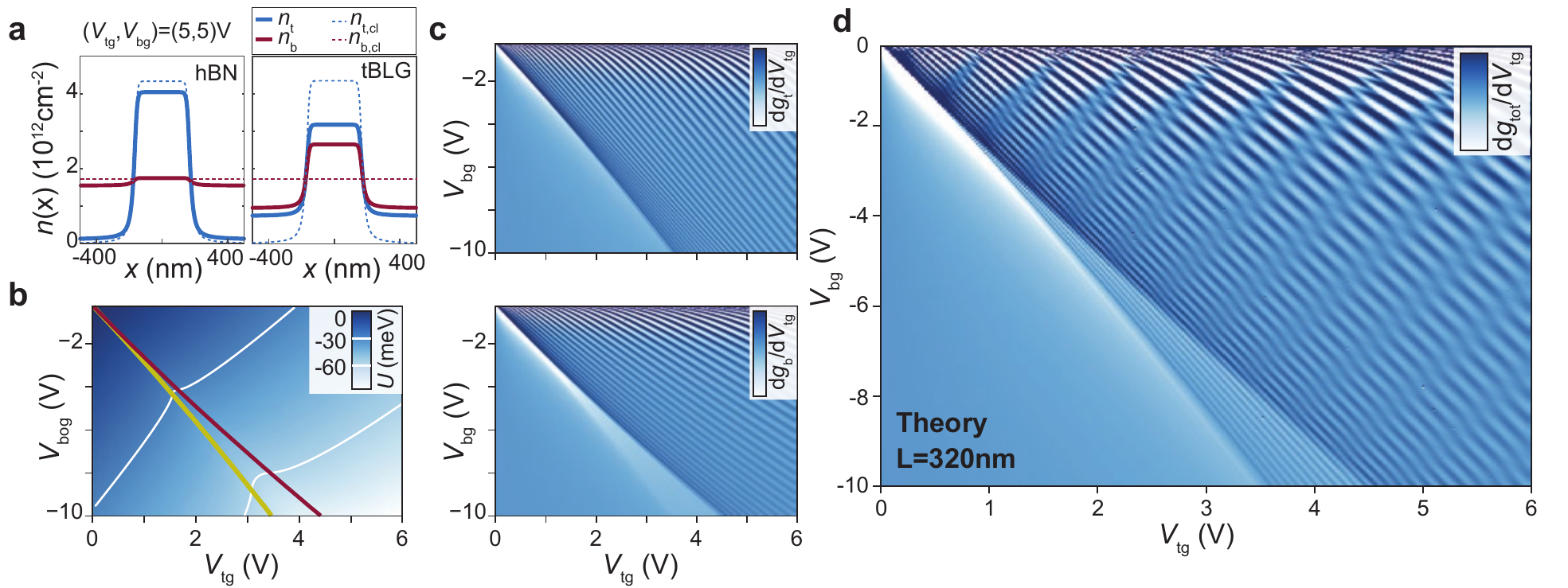}
	\caption{a) Numerically calculated layer density profile $\nT(x)$ (blue) and $\nB(x)$ (red), shaped by a top gate of size $\SI{320}{nm}$ and a global back gate for sample A with an hBN spacer (left) and sample B with a large-angle tBLG (right). The depicted gating condition is $(\VT,\VB)=(5,5)$V. The dashed lines show the classical result (neglecting the quantum capacitance). b) Interlayer energy difference $U(\VT,\VB)$. Zero density lines are marked with yellow and red lines. 
	c) Numerical derivative of the calculated normalized conductance, $\partial g/\partial \VT$.  Using the obtained $\nT$ and $\nB$, the conductance $g(\VT,\VB)$ of the top (top panel) and bottom layer (bottom panel) are calculated individually using a real-space Green's function approach. 
	d) The sum of the two differential conductances $\partial g_{\rm{tot}}/\partial \VT$ reproduces the experimental data in Fig. \ref{fig:FPdata}b.}
	\label{fig:theory}
\end{figure*}

We now compare the resonance pattern to tight-binding simulations. 
The underlying  density profiles $\nT(x)$ and $\nB(x)$  are obtained from a self-consistent electrostatic model where we assume that the dispersion relation remains linear, such that the carrier density formulas \cite{Liu2013} derived for single-layer graphene with quantum capacitance \cite{Luryi1988,Fang2008} taken into account can be readily applied. The extremely thin spacing between the two graphene layers leads to significant electrostatic coupling. Effectively, the channel potential of the top layer plays the role as a gate for the bottom layer, and vice versa.  For the twisted bilayer sample B (see Fig. \ref{fig:theory}a), the electrostatic coupling between the layers is significant, as can be seen by comparing to the classical density profiles (dashed lines).
In Fig. \ref{fig:theory}b we calculate the interlayer energy difference $U(\VT,\VB)$ for sample B. The maximum value we can reach is $U=\SI{80}{meV}$ in our device. We note here that the formula given in Refs.\cite{Sanchez-Yamagishi2017,Sanchez2012} for the displacement field (i.e. $D=1/2(\CT\VT-\CB\VB)$) only holds under the condition $\nT=-\nB$. Apparently, lines of constant $U$ (white lines in Fig. \ref{fig:theory}b) do not have a constant slope in the ($\VT,\VB$)-map. A more detailed comparison is given in the Supplemental Material.

In order to see whether the electrostatic model is in agreement with the experiment, we preform transport simulations  based on a real-space Green's function approach, considering two dual-gated, electronically decoupled graphene layers. To optimize the visibility of the Fabry-P\'erot interference fringes, we implement periodic boundary hoppings along the transverse dimension \cite{Liu2012}, equivalent to the assumption of infinitely wide graphene samples. This is justified since $W\gg L$ in our device. The normalized conductances $\gT(V_\mathrm{tg},V_\mathrm{bog})$  and $\gB(V_\mathrm{tg},V_\mathrm{bog})$ for the top and bottom graphene layers, respectively, are calculated using carrier density profiles $\nT(x)$ and $\nB(x)$. The numerical derivative of the results are shown in Fig. \ref{fig:theory}c. To compare with the measurement, we consider the numerical derivative $\partial g_\mathrm{tot}/\partial\VT$ of the sum $\gT+\gB=g_\mathrm{tot}$ (Fig. \ref{fig:theory}d). The excellent agreement to the measurement (Fig. \ref{fig:FPdata}b) is a strong indication that the wavefunctions of the top and bottom layer are essentially decoupled and individually tunable. 

In addition, the tight-binding theory allows us now to compare the electrostatic model to the experiment and to estimate the precision of the obtained value for the graphene interlayer capacitance $\CBLG$. 
For the cavity $L=\SI{320}{nm}$ and for $\VB=\SI{-10}{V}$ we observe $N=11\pm1$ modes between the two zero density lines in the experimental data (Fig. \ref{fig:FPdata}b) and $N=11\pm0.5$ modes in the tight-binding data (Fig. \ref{fig:theory}d). Since the splitting of zero-density lines is proportional to $\CBLG$ we estimate the error to be $\approx 10\%$ for $\CBLG$ and therefore we estimate the dielectric thickness of graphene $\dBLG=2.6\pm\SI{0.2}{\angstrom}$.

\paragraph{Conclusion}.
We have performed transport experiments for two representative cases of decoupled layers of graphene. We investigated two devices, one where decoupling is achieved by a thin hBN layer (sample A) and the other where the decoupling is given by the large momentum mismatch between graphene layers due to a large twist angle (sample B). In both cases we observed a clear splitting of the charge neutrality points in a two-terminal measurement with the strength of the splitting given by $C_\mathrm{q}/\CBLG$. By comparing to a self-consistent electrostatic model we extracted a very large geometric interlayer capacitance $\CBLG=7.5\pm\SI{0.7}{\mu Fcm^{-2}}$ for the twisted bilayer graphene sample, which we explained by taking into account an effective dielectric thickness of graphene of $\dBLG=2.6\pm\SI{0.2}{\angstrom}$. 
In a further step, we investigated Fabry-Pérot fringes that originate from p-n-p cavities created with a local top- and a global back gate. We were able to form a p-n-p cavity in only one of the layers and could tune the wavelength in each layer individually. In an $L=\SI{320}{nm}$ cavity we observed the first mode in the bottom layer, while we had already filled  ten modes in the top layer. The measurements are in very good agreement with the results from tight-binding simulations based on two graphene layers electronically decoupled but electrostatically coupled through their quantum capacitances.
Our work emphasizes that the finite thickness of 2D materials is relevant for the electronic properties of Van-der-Waals heterostructures where conducting layers are in close proximity. %Additionally, the Fabry-Pérot experiment highlights the importance of the layer degree of freedom for quantum transport experiments and may inspire quantum information processing schemes that depend on the layer quantum number.

\section*{Acknowledgements}
We acknowledge financial support from the European Graphene Flagship, the Swiss National Science Foundation via NCCR Quantum Science and Technology and from the Deutsche Forschungsgemeinschaft through SFB 1277, project A07 and the Taiwan Minister of Science and Technology (MOST) under Grant No. 107-2112-M-006-004-MY3. The work is also supported  by  the  National  Science Center under the contract DEC-2018/29/B/ST3/01892, and in part by PAAD Infrastructure co-financed by Operational Programme Innovative Economy, Objective 2.3. Growth of hexagonal boron nitride crystals was supported by the Elemental Strategy Initiative conducted by MEXT, Japan and the CREST (JPMJCR15F3), JST.

%\bibliographystyle{apsrev4-1}
%\begingroup
%\inputencoding{latin1}


%\endgroup
\clearpage
\setcounter{figure}{0}
\renewcommand{\thefigure}{S\arabic{figure}}
\renewcommand{\theequation}{S\arabic{equation}}

\appendix

\part{Supplemental Material}

\section{Accuracy of the fitting procedure}\label{app:accuracy}
\begin{figure*}
	\centering
	\includegraphics[width=1\textwidth]{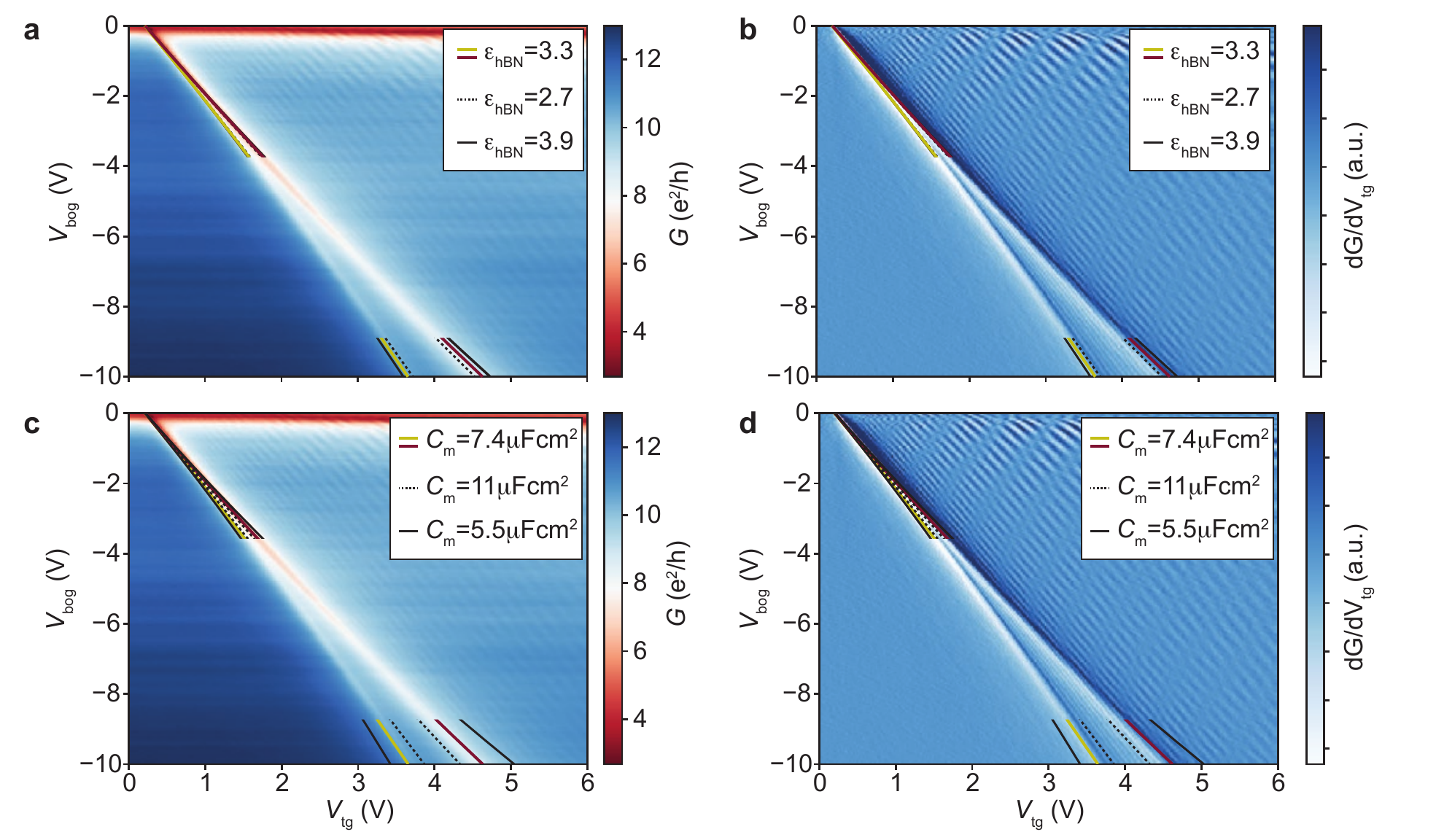}
	\caption{a), Conductance and b) numerical derivative for different values of the hBN dielectric constant $\epsilon_\mathrm{hBN}$. In the main text we use$\epsilon_\mathrm{hBN}=3.3$.
	In c) and d) we show that the splitting is highly sensitive to the interlayer distance $\dBLG$, i.e. for values of $\dBLG=\SI{0.8}{\angstrom}$ (dashed line) or $\dBLG=\SI{1.6}{\angstrom}$ (solid black line) the splitting is clearly under or overestimated, respectively.
} 
	\label{fig:suppfig2fitting}
\end{figure*}

In the main text we argue that we are able to determine the interlayer capacitance $\CBLG$  accurately. We determine $\CBLG$ by comparing the measured splitting of the charge neutrality points to the results we obtain in our capacitance model (Fig. 2c). The thickness of the top and bottom hBN layers were determined by AFM (atomic force microscopy) measurements with an accuracy of $\pm 0.5nm$. The ratio $\CT/\CB$ determines the slope in the $(\VT,\VB)$ map and is in agreement with the ratio of $\dT/\dB$ obtained by AFM. The dielectric constant of hBN is not precisely known in our case (the error can be as large as $20\%$), however this does not crucially influence our analysis as we show in Fig.\ref{fig:suppfig2fitting}a and b where we depict zero-density lines for $\epsilon_{\mathrm{hBN}}=2.7, 3.3,3.9$.

The Fabry-Pérot resonator serves as an excellent tool to determine the accuracy of the electrostatic model and with this the accuracy of the value that we extract for $\CBLG$. For the cavity $L=\SI{320}{nm}$ and for $\VB=\SI{-10}{V}$ we observe $N=11\pm1$ modes between the two zero density lines in the experimental data (Fig. 3b) and $N=11\pm0.5$ modes in the tight-binding data (Fig 4c.) based on the electrostatic model. Since the splitting of the lines is proportional to $\CBLG$ we can estimate the systematic error to be $\approx 10\%$ and therefore $\CBLG=7.4\pm\SI{0.7}{\mu F/cm^2}$ which translates into an error for the dielectric thickness of graphene of
$\dBLG=2.6\pm\SI{0.2}{\angstrom}$.

\section{Comparison to the elctrostatic model in Refs.\cite{Sanchez-Yamagishi2017,Sanchez2012}}

\begin{figure*}
	\centering
	\includegraphics[width=1\textwidth]{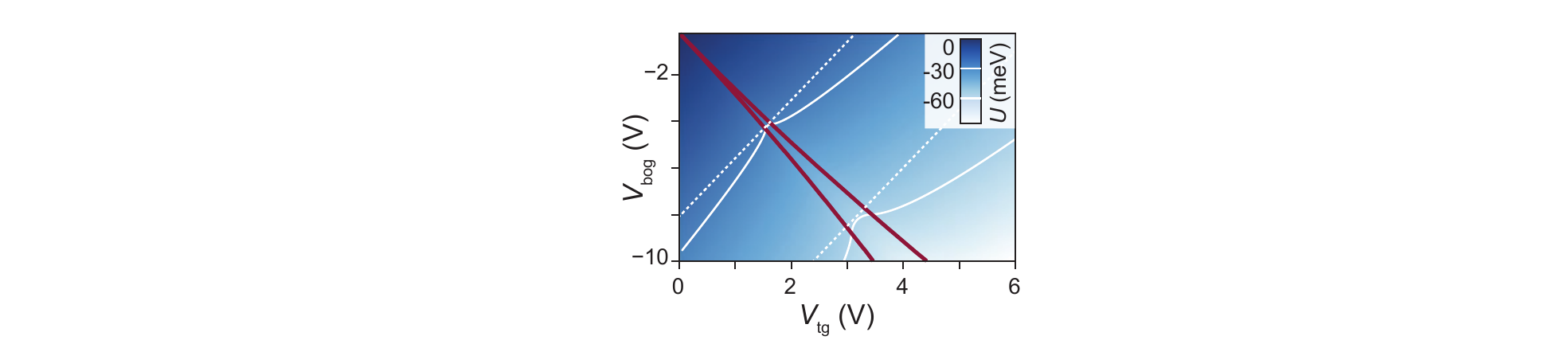}
	\caption{a), Interlayer energy difference $U$ calculated by our iterative electrostatic model. red lines denote zero-density in the upper/lower layer. Along the solid white lines, $U=(30,60)$meV. The white dashed lines correspond to the same values of $U$ but calculated using the formula in Refs. \cite{Sanchez2012,Sanchez-Yamagishi2017}.} 
	\label{fig:suppfig1U}
\end{figure*}

In Refs. \cite{Sanchez-Yamagishi2017,Sanchez2012}, the displacement field between the graphene layers is calculated using:
\begin{equation}\label{eq:D}
D=\frac{1}{2}\left(\CT\VT-\CB\VB \right)
\end{equation}
This is not in agreement with the interlayer energy $U=D\cdot d/\epsilon_0$ that we calculate using an itterative model to obtain the layer carrier densities \cite{Liu2013} (see Fig.\ref{fig:theory}a). For a more detailed comparison, we plot lines of constant $U$ using our model and the one of Refs. \cite{Sanchez-Yamagishi2017,Sanchez2012}. We rephrase equation \ref{eq:D}:
\begin{equation}\label{eq:D}
\left.\VT(\VB)\right|_{U=\mathrm{const.}}=\frac{\CT}{\CB}\VT-\frac{2U\epsilon_0}{d\CB}
\end{equation}
The solid lines in Fig.\ref{fig:suppfig1U} are lines of constant $U$ for the values $U=\SI{-30}{meV}$ and $U=\SI{-60}{meV}$, the dashed lines are calculated with the simplified formula (eq. \ref{eq:D}) for the same values of $U$. The models agree for $n_{\rm{tot}}=0$, i.e. if $\nT=-\nB$.

\section{Comparison to previous studies}

%\begin{tabularx}{\linewidth}{@{}>{\bfseries}l@{\hspace{1em}}X|@{\hspace{1em}}X|@{\hspace{1em}}X|@{\hspace{1em}}X|@{\hspace{1em}}X|}
\begin{tabularx}{\linewidth}{|l||X|X|X|X|X|}
	\hline 
	&	Schmidt 2008 \cite{Schmidt2008} & Lucian 2011\cite{Lucian2011} & Sanchez-Yamagishi 2012 \cite{Sanchez2012} & Sanchez-Yamagishi 2017 \cite{Sanchez-Yamagishi2017} & This work \\ 
	\hline\hline
		Technique& SdH measured with top/back gate for tBLG on SiO2 & STM  spectroscopy (also at large $\theta$)  &  SdH at $B=\SI{4}{T}$& QHE in 2-terminal geometry & Fabry-Pérot spectroscopy \\ 
	\hline 
	Observation& Tunability of layer density  & $v_{\rm{F}}$ as a function of twist, decoupling for large $\theta>\SI{20}{\degree}$  & SdH originate from layers, can be tuned by $U$, allow to extract $\CBLG$ & Counterpropagating $\nu=\pm1$ QH states exhibit plateaus, but not the $\nu=\pm2$ states due to backscattering & Splitting of CNPs and layer dependent resonances \\  
	\hline 
	$B_{\rm{min}}$ for SdH &$<\SI{3}{T}$   & & $1-\SI{2}{T}$   & $1-\SI{2}{T}$ & $\approx\SI{50}{mT}$ \\ 
	\hline 
	Lengthscale & $l_{\rm{B}}=\SI{15}{nm}$ & STM tip $\sim$nm & $l_{\rm{B}}=\SI{12}{nm}$ & $l_{\rm{B}}=\SI{12}{nm}$ & $\lambda=\SI{600}{nm}$ \\ 
	\hline 
	$\CBLG$ and $d$ & $d=\SI{1.2}{nm}$ (?) & &  $\CBLG=\SI{7.5}{\mu Fcm^{-2}}$  and $d=\SI{1.2}{\angstrom}$ & same sample as 2012? & $\CBLG=\SI{7.5}{\mu Fcm^{-2}}$ and $d=\SI{1.2}{\angstrom}$ \\ 
	\hline 
\end{tabularx}

\bigskip
Previous studies were done at high magnetic field, i.e. coupling or decoupling of quantum hall edge states has been investigated. Except for the STM study, the bulk was not probed. In our case we probe the bulk. The corresponding lengthscale of the objects which are decoupled is an order of magnitude larger in our case. This long-wavelength regime could not be reached in previous studies. Demonstrating decoupling at $B={0}{T}$ and over large areas is crucial if one intends to exploint the ultrahigh capacitance of the parallel plate capacitor for detection, capacitive coupling or even energy storage.
\\

\section{First principle calculations}

\begin{figure}[htp]
	\centering
	\includegraphics[width=0.7\columnwidth]{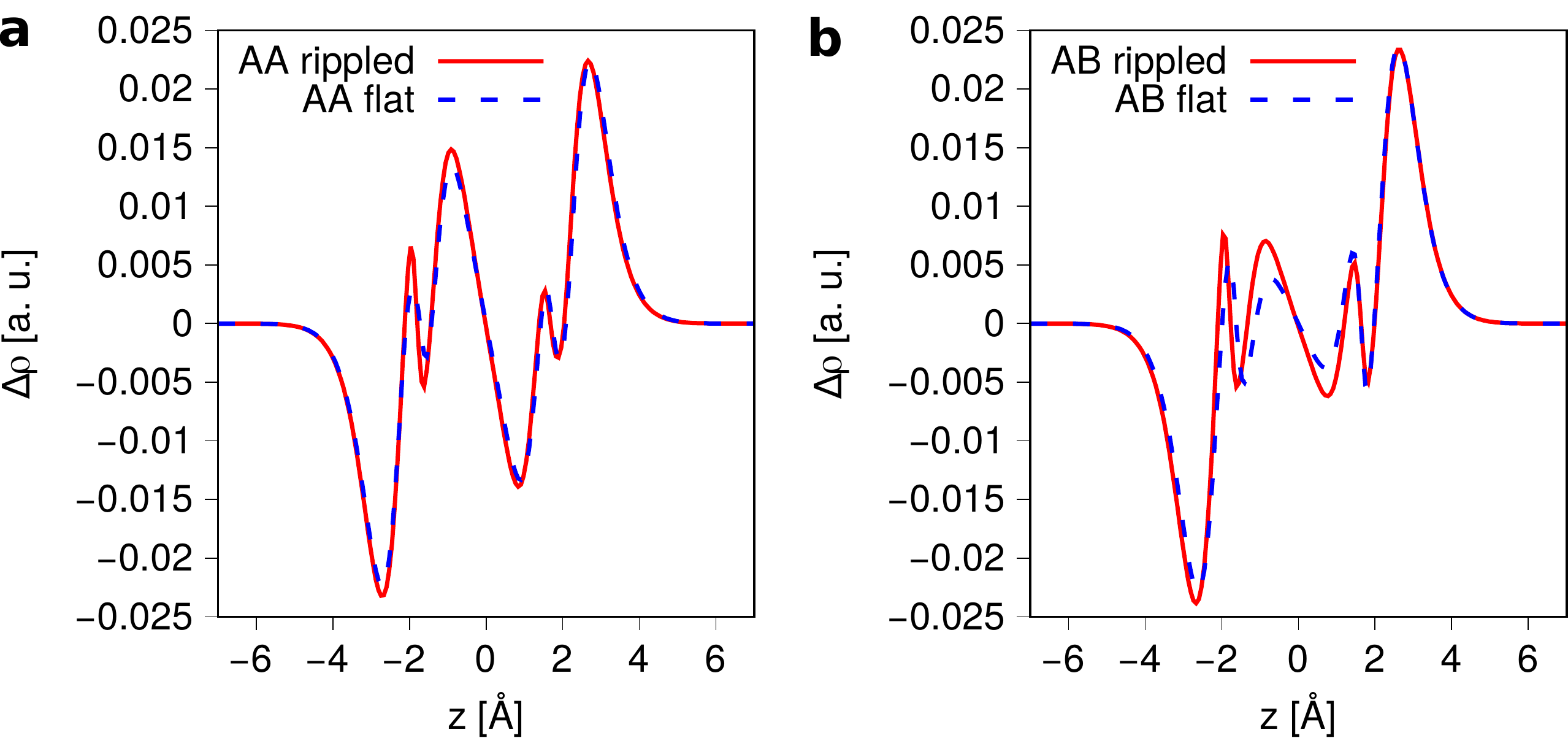}
	\caption{Comparison of the induced charge density profile $\Delta \rho(z)$ for flat and rippled AA (a) and AB (b) bilayer graphene. The rippling of size $\Delta d_z=0.2\,$\AA was introduced in the bottom layer of BLGs. \label{fig:rippling}}
\end{figure}

First principles calculations have been performed using \textsc{Quantum Espresso} \cite{QE2009,QE2017} package. In all calculations the ultrasoft pseudopotential \cite{rkjjus} with the Perdew-Burke-Ernzerhof \cite{pbe} implementation of exchange-correlation functional was used, with the kinetic energy cut-offs for the wave function and charge density 48\,Ry and 480\,Ry respectively. We used lattice constant of graphene $a=2.46\, \text{\AA}$, the same as in the electrostatic model.
To avoid spurious interactions between periodic copies of the system a vacuum of $22\,\text{\AA}$ was introduced.  Calculations with non-zero external electric field were done with the enabled dipole correction \cite{Bengtsson1999}. 
%in all cases.  
Self-consistency has been achieved for $30\times 30$ Monkhorst-Pack\cite{monkhorst} $k$-point grid. 
%in case of graphene and AA and AB stacked bilayer graphene. 
Calculations of the integrated local density of states (ILDOS) and charge density profiles were performed with a dense, $120\times 120$ Monkhorst-Pack $k$-point grid.

Calculations of the twisted bilayer graphene were performed with  the $k$-point grids $3 \times 3$ and $21 \times 21$ for self-consistent and non-self-consistent calculations respectively. The dispersion correction due to the Van der Waals interaction in bilayer graphenes was taken into account via DTF-D3 \cite{dftd3} method. Initial atomic positions in bilayer graphene structures were optimized using quasi--Newton scheme, as implemented in \textsc{Quantum Espresso} \cite{QE2009,QE2017}. The relaxation of internal forces acting on atoms was done with the force and energy convergence thresholds $10^{-3}$\, Ry/bohr and $10^{-4}\,$ Ry/bohr respectively.

\begin{figure}[h]
	\centering
	\includegraphics[width=0.6\columnwidth]{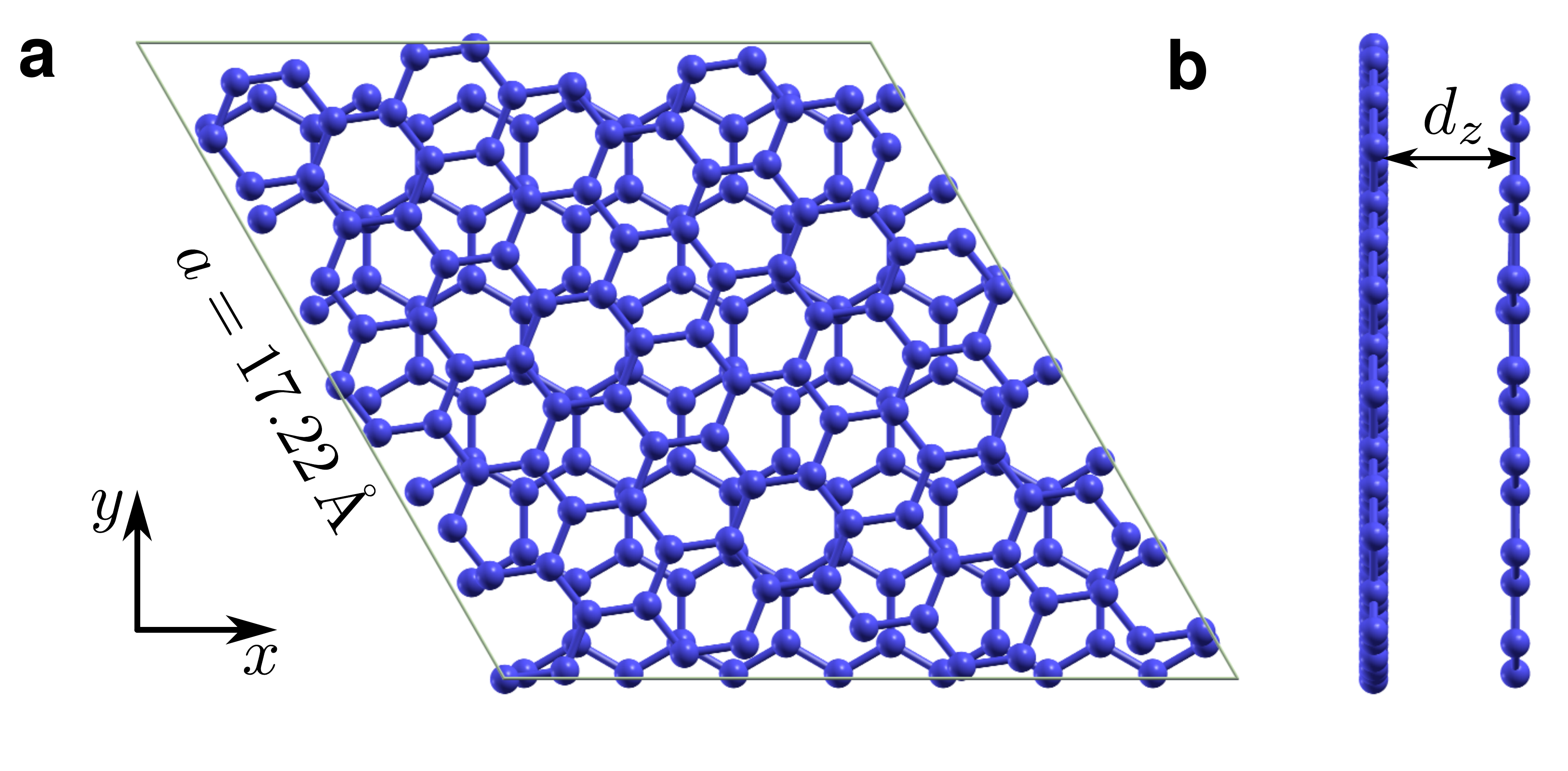}
	\caption{Schematics of the crystalline structure of the unit cell of $22^\circ$--twisted bilayer graphene with 196 carbon atoms. a) Top view with the indicated lattice constant $a=17.22\,\AA = 7a_{grp}$,~$a_{grp}=2.46\,\AA$ is the lattice constant of graphene.
		b) side view with indicated interlayer lattice distance $d_z$.
		\label{fig:blg_super}}
\end{figure}

In Fig. \ref{fig:blg_super} we show the computational unit cell. For the twist angle 22$^\circ$ tBLG the unit cell is an approximate unit cell due to a tiny incommensurability of the top and bottom layers. The relaxed structure displays a small rippling in only one layer whereas the the second remains generally flat. 
In effect, the interlayer distance $d_z$ vary from 3.42\,\AA\, to 3.6\,\AA. We have checked to what extent the out-of-plane lattice distortions can modify the induced charge density $\Delta \rho(z)$. In Fig. \ref{fig:rippling} we plot $\Delta \rho(z)$ 
for flat and rippled AA and AB BLG structures. Similarly to tBLG  rippling of size $\Delta d_z = 0.2\,$\AA was introduced only in the bottom layer of BLG. One can see, that lattice distortions amplify $\Delta \rho(z)$ in the interlayer region. Therefore, we conclude that the flattening of $\Delta \rho(z)$ in the interlayer region is an intrinsic feature of tBLG.

\begin{figure}[h]
	\centering
	\includegraphics[width=0.7\columnwidth]{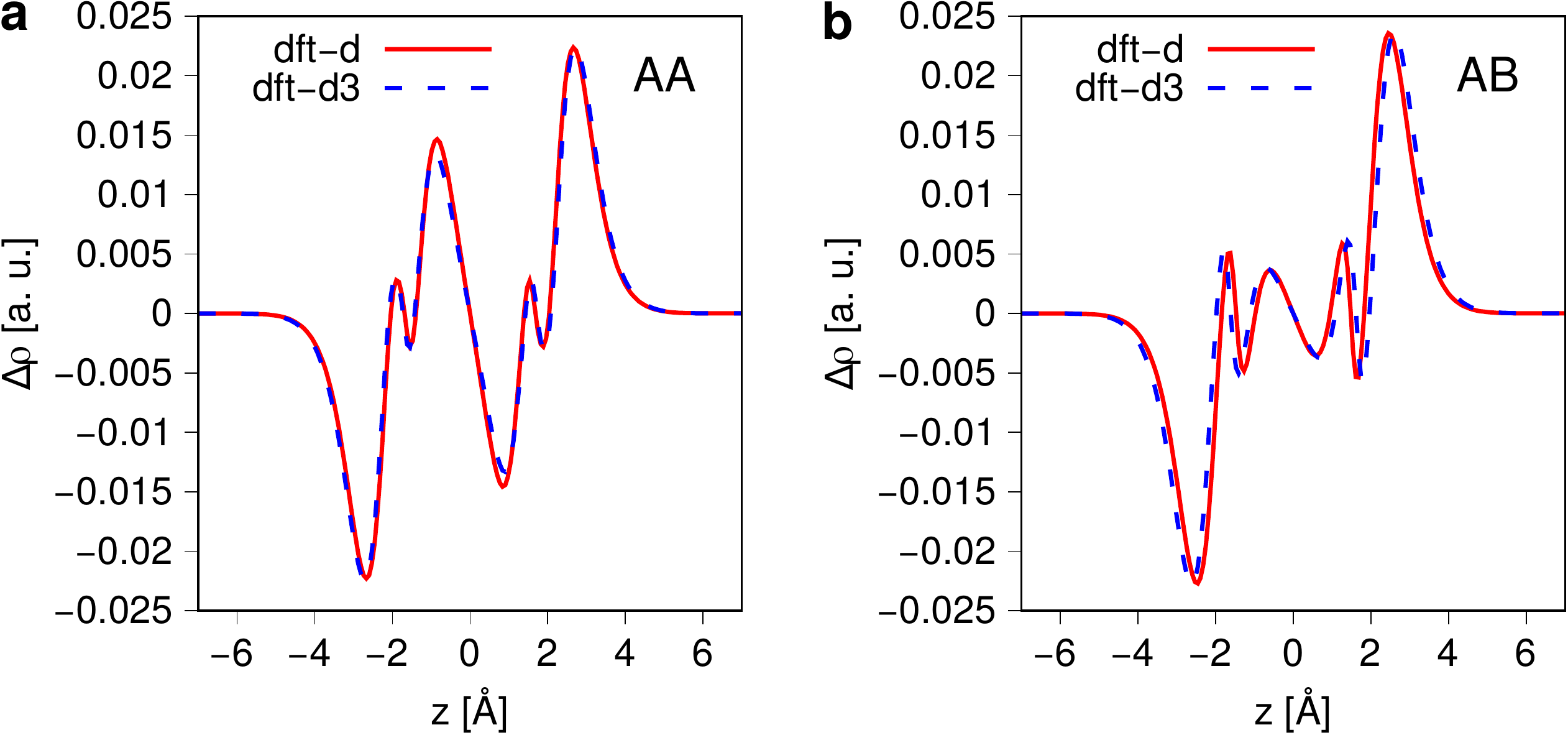}
	\caption{Comparison of the induced charge density $\Delta \rho(z)$ for AA (a) and AB BLG (b) for DFT-D and DFT-D3 types of dispersion correction. The interlayer distances are collected in Table \ref{inter_distance}.
		\label{fig:cdens_vdv_comparison}}
\end{figure}

In Fig. \ref{fig:cdens_vdv_comparison} we show the induced charge densities $\Delta \rho(z)$ for AA and AB BLG and for two different types of dispersion corrections, DFT-D \cite{dftd_1,dftd_2} and DFT-D3 \cite{dftd3}. For AB stacking the optimized interlayer distances for DFT-D  and DFT-D3 differ by 0.27\, AA. For AA BLG this difference is $0.1\,$\AA (see Table \ref{inter_distance}). It is seen, that $\Delta \rho(z)$ is independent of the type of the chosen dispersion correction method and of small differences in the interlayer distances.

\begin{table}[h]
	\caption{\label{inter_distance} Calculated interlayer distances for AA and AB BLGs for different types of dispersion corrections.}
	\begin{tabular}{|c|c |c| c|} 
		\hline
		AA DFT-D & AA DFT-D3 & AB DFT-D & AB DFT-D3 \\ [0.5ex] 
		\hline
		3.536\,\AA & 3.64 \,\AA& 3.24\,\AA & 3.51\,\AA \\ 
		\hline
	\end{tabular}
\end{table}

\section{Electrostatic model}\label{sec:electrostatics}
A schematic of the model for the bilayer graphene structure introducing the relevant quantities is shown below. The horizontal axis is the $z$-direction, the vertical axis denotes energy for electrons.

\begin{figure}
	\centering
	\includegraphics[width = 1\textwidth]{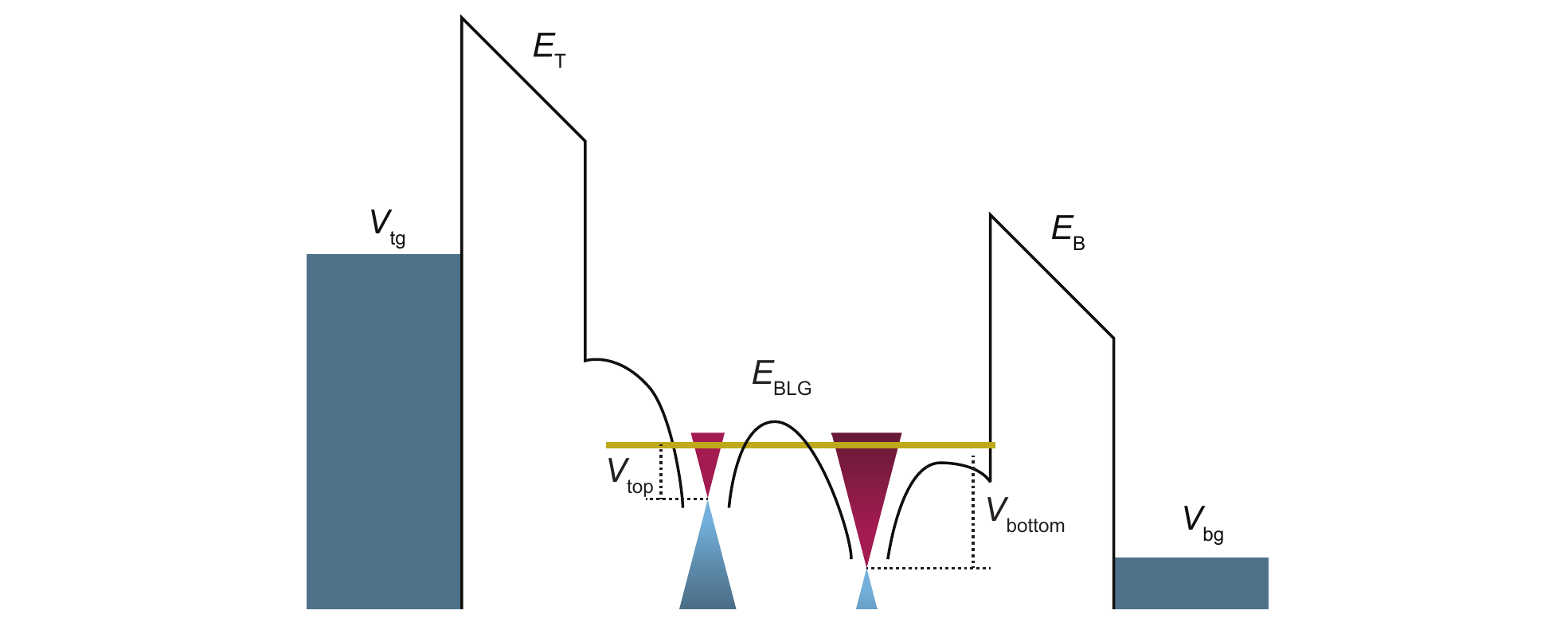}	
	\caption{Schematic model for gated, decoupled bilayer graphene}
\end{figure}

\paragraph{Basis of the Electrostatics.}
Taking a cylindrical volume with axis in $z$-direction, and assuming a homogeneous field $\vec{E}=(0,0,E)$ as to be expected in the structure sketched above, Poisson's equation simplifies to
	\[ E_\mathrm{R} -E_\mathrm{L} = \frac{\sigma_\mathrm{encl}}{\epsilon_0},\]
where $E_\mathrm{L}$ is the $z$-component of the electric field on the left cylinder surface, $E_\mathrm{R}$ is the electric field on the right cylinder surface, and $\sigma_\mathrm{encl}$ is the encosed two-dimensional charge density.
For the following arguments we will assume that the two graphene layers can be modeled as two thin plates that are coupled by the capacitance that we determine from the measurements, $\CBLG$. 
For simplicity we first assume that this capacitor exhibits a constant electric field $\EBLG$ and dielectric constant $\epsBLG=\epsilon_0$ and that the plates are separated by $\dBLG$. 
\[ \CBLG = \frac{\epsilon_0}{\dBLG}\]
Only later (in the main text) we will argue that $\CBLG$ contains both, the capacitance of a single graphene sheet (with $\epsilon=6.9$) and the true interlayer capacitance (with $\epsilon=1$) which are connected in series.

\paragraph{Energy considerations.} We will now relate the different chemical potentials in the structure, keeping a sense for the symmetry of the device in $z$-direction. Assuming that the electric fields inside the top and bottom gates are zero, we find that the charge in the top layer $\sigmaT=e\nT$ is enclosed by the electric fields of the top graphene sheet to the topgate, $\ET$ and the electric field between the graphene sheets, $\EBLG$. Therefore: 
\begin{eqnarray}
\EBLG-\ET=\frac{e\nT}{\epsilon_0}\\
\EB-\EBLG=\frac{e\nB}{\epsilon_0}
\end{eqnarray}
The electric fields are related to the potential differences between the layers:
\begin{eqnarray}
\ET=(\VT-\Vtop)\epsilonhBN/\dT\\
\EBLG=(\Vtop-\Vbottom)/\dBLG\\
\EB=(\Vbottom-\VB)\epsilonhBN/\dB
\end{eqnarray}
Here, $\dT$ and $\dB$ are the thicknesses of the respective top and bottom layers of hBN and $\dBLG$ is the distance between the graphene sheets. Now we realize that $e\Vtop=\EFT$ and $e\Vbottom=\EFB$. We then combine the above equations to obtain equations that relate the top- and bottomgate voltages to the densities:
\begin{eqnarray}
\frac{1}{e\dBLG}(\EFT-\EFB) - \frac{\epsilonhBN}{e\dT}(e\VT-\EFT) = \frac{e\nT}{\epsilon_0}   \label{eq:Vtgvsn}\\
\frac{\epsilonhBN}{e\dB}(\EFB-e\VB)- \frac{1}{e\dBLG}(\EFT-\EFB) = \frac{e\nB}{\epsilon_0}    \label{eq:Vbgvsn}
\end{eqnarray}

These equations allow us to determine lines of constant density. For this purpose, we take a partial derivative of each of the two equations with respect to $\nT$ at constant $\nB$, and vice versa.  We customize the equations using the capacitances e.g. $\CBLG=\epsilon_0/\dBLG$ and  realize that
\[ \frac{\partial \nT}{\partial\EFT} = \DOST,\quad\mbox{and}\quad \frac{\partial \nB}{\partial\EFB} = \DOSB \]
represent the densities of states at the Fermi energy of the two graphene layers.

This results in the four equations
\begin{eqnarray}
\frac{1}{e}\left.\frac{\partial\VT}{\partial \nT}\right|_{\nB=\const}  & = & \frac{\dT}{\CBLG\dBLG\epsilonhBN} + \frac{\dT}{\dBLG\epsilonhBN}\frac{1}{e^2\DOST}  + \frac{1}{e^2\DOST}\label{toplayer4}\\
\frac{1}{e}\left.\frac{\partial\VB}{\partial \nT}\right|_{\nB=\const}  & = & -\frac{\dB}{\dBLG\epsilonhBN}\frac{1}{e^2\DOST} \label{botlayer4}\\
\frac{1}{e}\left.\frac{\partial\VT}{\partial \nB}\right|_{\nT=\const}  & = & -\frac{\dT}{\dBLG\epsilonhBN}\frac{1}{e^2\DOSB} \label{toplayer4a}\\
\frac{1}{e}\left.\frac{\partial\VB}{\partial \nB}\right|_{\nT=\const} & = & \frac{\dB}{\CBLG\dBLG\epsilonhBN} + \frac{\dB}{\dBLG\epsilonhBN}\frac{1}{e^2\DOSB} + \frac{1}{e^2\DOSB} \label{botlayer4a}
\end{eqnarray}
These equations resemble the symmetry of the sample. Replacing all indices $top$ with indices $bottom$, and exchanging $\nT$ and $\nB$ in the first two equations results in the second two equations.

Taking the ratios of the two eqs.~\eqref{toplayer4} and \eqref{botlayer4}, and also of \eqref{botlayer4a} and \eqref{toplayer4a} gives two equations for the slopes of constant density lines in the $\VT$-$\VB$ plane, i.e., the plane of the measurement
\begin{eqnarray}
\left.\frac{\partial\VT}{\partial\VB}\right|_{\nB=\const} & = & -\frac{\CB}{\CT}\left(1+\frac{ e^2\DOST}{\CBLG}\right) - \frac{\dBLG\epsilonhBN}{\dB} \\
\left.\frac{\partial\VB}{\partial\VT}\right|_{\nT=\const} & = & -\frac{\CT}{\CB}\left(1 + \frac{e^2\DOSB}{\CBLG}\right) - \frac{\dBLG\epsilonhBN}{\dT}.
\end{eqnarray}

In particular for the case of tBLG, the terms $\frac{\epsilonhBN\dBLG}{\dB}$ and $\frac{\epsilonhBN\dBLG}{\dT}$ are small and the zero density lines simplify to
\begin{eqnarray}
\left.\frac{\partial\VT}{\partial\VB}\right|_{\nB=\const} & \approx & -\frac{\CB}{\CT}\left(1+\frac{ e^2\DOST}{\CBLG}\right) 
 \label{main1}\\
\left.\frac{\partial\VB}{\partial\VT}\right|_{\nT=\const} & \approx & -\frac{\CT}{\CB}\left(1 + \frac{e^2\DOSB}{\CBLG}\right) 
\label{main2}
\end{eqnarray}
These two equations are the main result of the presented calculation and are discussed in the main text.

\bigskip
\paragraph{Discussion of the result.}
If the two voltages $\VT$ and $\VB$ are tuned such that both layers are at charge neutrality, for twisted bilayer graphene we expect that
\[ \frac{ e^2\DOST}{\CBLG} \approx 0,\quad\mbox{and}\quad \frac{e^2\DOSB}{\CBLG}\approx 0. \]
We therefore have approximately
\begin{eqnarray}
\left.\frac{\partial\VT}{\partial\VB}\right|_{\nB=\const} & = & -\frac{\CB}{\CT} - \frac{\dBLG\epsilonhBN}{\dB} \\
\left.\frac{\partial\VB}{\partial\VT}\right|_{\nT=\const} & = & -\frac{\CT}{\CB} - \frac{\dBLG\epsilonhBN}{\dT}.
\end{eqnarray}
This means that the observed charge-neutrality lines for the two layers do not have exactly the same slope at their intersection. However, since the ratios $\dBLG\epsilonhBN/\dT$ and $\dBLG\epsilonhBN/\dB$ can be expected to be rather small for large-angle twisted bilayer graphene, the difference in slope may be hard to observe in this case.
In samples, where a hBN layer separates the two graphene layers, it is not obvious, if the above approximation still holds. As a result, the different slopes of the two charge neutrality lines should tend to be more pronounced.
We also see in eqs.~\eqref{main1} and \eqref{main2} that the deviations from linearity of the constant density lines is governed entirely by the densities of states of the two layers.

\bigskip
\paragraph{Explicit formula for the zero density lines}

To find an expression for the zero-density line in the $(\VT,\VB)$-plane, we insert the Fermi-energy into equations \ref{eq:Vbgvsn} and \ref{eq:Vtgvsn}.
%\begin{eqnarray}
%\frac{\hbar\vF\sqrt{\pi}}{e\dBLG}(\sqrt{\nT}-\sqrt{\nB}) - %\frac{\epsilonhBN}{e\dT}(e\VT-\hbar\vF\sqrt{\nT\pi}) = \frac{e\nT}{\epsilon_0} \\
%\frac{\epsilonhBN}{e\dB}(\hbar\vF\sqrt{\nB\pi}-e\VB)- %\frac{\hbar\vF\sqrt{\pi}}{e\dBLG}(\sqrt{\nT}-\sqrt{\nB}) = \frac{e\nB}{\epsilon_0}    
%\end{eqnarray}
For the zero-density line of the top-layer, $\nT=0$ and therefore:
\begin{eqnarray}
\frac{\hbar\vF\sqrt{\pi}}{e\dBLG}\sqrt{\nB} - \frac{\VT\epsilonhBN}{\dT} = 0 \\
\frac{\epsilonhBN}{e\dB}(\hbar\vF\sqrt{\nB\pi}-e\VB)- \frac{\hbar\vF\sqrt{\pi}}{e\dBLG}(\sqrt{\nB}) = \frac{e\nB}{\epsilon_0}    
\end{eqnarray}
Inserting $\nB$ from the first equation
%\begin{eqnarray}
%\sqrt{\nB} = \left(e\VT\frac{\dBLG}{\epsilonhBN\dT}\frac{1}{\hbar\vF\sqrt{\pi}}\right)
%\end{eqnarray}
into the second equation and solving for $\VB$, %assuming $\VT<0$
%\begin{eqnarray}
%\frac{1}{e\epsilonhBN\dB}(e\VT\frac{\dBLG}{\epsilonhBN\dT}-e\VB)- \frac{1}{e\dBLG}(e\VT\frac{\dBLG}{\epsilonhBN\dT}) = \pm\left(e\VT\frac{\dBLG}{\epsilonhBN\dT}\frac{1}{\hbar\vF\sqrt{\pi}}\right)^2\frac{e}{\epsilon_0}    
%\end{eqnarray}
%above eq to be deleted
%\begin{eqnarray}
%\VT\frac{\dBLG}{\dT}-\VB-\VT\frac{\dB}{\dT} = \pm\left(e\VT\frac{\dBLG}{\dT}\frac{1}{\hbar\vF\sqrt{\pi}}\right)^2\frac{e}{\epsilon_0} \dB   
%\end{eqnarray}
%above eq to be deleted
%\begin{eqnarray}
%\VB=\VT(\frac{\dBLG}{\epsilonhBN\dT}-\frac{\dB}{\dT}) \mp\left(e\VT\frac{\dBLG}{\epsilonhBN\dT}\frac{1}{\hbar\vF\sqrt{\pi}}\right)^2\frac{e}{\epsilon_0} \epsilonhBN\dB   
%\end{eqnarray}
%above eq to be deleted
\begin{eqnarray}
\VB=(\frac{\dBLG\epsilonhBN}{\dT} - \frac{\dB}{\dT})\VT -\frac{e^3}{\hbar^2\vF^2\pi}\frac{\dBLG^2\dB\epsilonhBN}{\dT^2\epsilon_0}\sign(\VT)\VT^2
%\approx -\frac{\dB}{\dT}\VT -\frac{e^3}{\hbar^2\vF^2\pi}\frac{\dBLG^2\dB\epsilonhBN}{\dT^2\epsilon_0}\VT^2
\end{eqnarray}
Expressed in terms of capacitances:
\begin{eqnarray}
\VB (\VT)|_{\nT=0}  =  - \frac{\CT}{\CB}\VT \left(1 +\frac{e^3}{\hbar^2\vF^2\pi}\frac{\CT}{\CBLG^2} \sign(\VT)\VT \right)+\frac{\CT}{\CBLG}\VT
\end{eqnarray}
Again, the last term is small. For the zero-density line of the bottom layer we find:
\begin{eqnarray}
\VB (\VT)|_{\nB=0}  = \sign{\VT}\frac{\CBLG}{2\CB}\frac{\hbar\vF\sqrt{\pi}}{e^3} \left((\CBLG+\CT)\hbar\vF\sqrt{\pi}  -  \sqrt{(\CBLG+\CT)^2\hbar^2\vF^2\pi + \sign(\VT)4e^3\CT\VB}\right)
\end{eqnarray}

\end{document}